\documentclass[useAMS,usenatbib]{mn2e}
\usepackage{epsfig}
\usepackage{graphicx}
\usepackage{psfrag}
\usepackage{amssymb,amsmath}

\title[Calibrating the BAO scale using the CMB]{Lifting the degeneracy between geometric and dynamic distortions using the sound horizon from the CMB}
\author[A. J. Hawken, F. B. Abdalla,  G. H\"utsi, O. Lahav]{Adam J. Hawken$^{1}$, Fillipe B. Abdalla$^{1}$,  Gert H\"utsi$^{2}$$^{3}$, Ofer Lahav$^{1}$
\thanks{E-mail: ahawken@star.ucl.ac.uk; ghutsi@aai.ee; fba@star.ucl.ac.uk; lahav@star.ucl.ac.uk}\\
$^{1}$Department of Physics and Astronomy, University College London, Gower Street, London\\
$^{2}$Tartu Observatory, Toravere 61602, Estonia\\
$^{3}$Max-Planck Institute fur Astrophysick, Karl-Schwarzchild-Str. 1, D-85741 Garching, Germany}

\begin{document}

\pagerange{\pageref{firstpage}--\pageref{lastpage}} \pubyear{2011}
\maketitle

\label{firstpage}

\begin{abstract}
The degeneracy between geometric (Alcock-Paczynski) and dynamic (redshift space) distortions in the pattern of the galaxy distribution has been a long standing problem in the study of the large scale structure of the universe. We examine the possibility of lifting this degeneracy and constraining cosmological parameters by using the Baryon Acoustic Oscillation (BAO) scale as a feature of known physical size, the sound horizon $r_{\rm s} \approx 150 {\rm Mpc}$. We callibrate this scale with the equivalent feature in the Cosmic Microwave Background (CMB). First, we construct a toy model of a power spectrum which includes the BAO as well as geometric and dynamic distortions. By adding a prior onto the sound horizon of $\sim 1 \%$ we show, using a Fisher matrix analysis, that error ellipses for line of sight and tangential distortion parameters shrink by a factor of two for a $20(h^{-1}{\rm Gpc})^3$ `DESpec/BigBOSS'-like galaxy survey including shot noise. This improvement is even more marked in smaller surveys. We also carry out a Monte Carlo Nested Sampling analysis on our parameter space. We find that Monte Carlo and Fisher methods can agree reasonably well for surveys with large volume but differ greatly for small volume surveys.
\end{abstract}


\section{Introduction}\label{Introduction}

In order to derive knowledge of cosmology from observations of the large scale structure of the universe, we must accurately record the angular positions and distances to billions of galaxies and then analyse their statistical distribution. To determine distances in cosmology it is first necessary to have a fiducial cosmological model that allows the observer to calculate quantities such as the angular diameter distance $d_A(z)$ and Hubble function $H(z)$. Inevitably the underlying model the cosmologist has is incorrect in some way. This results in geometric distortions known as the Alcock-Paczynski effect \citep{1979Natur.281..358A} - distances measured along the line of sight look different to those measured perpendicular to the line of sight \citep{2005MNRAS.364..743N, 2011MNRAS.tmp.1599B, 2011arXiv1105.2037K}.

The redshift of a galaxy, $z$, is not a true measure of a galaxy's distance but a measure of the recessional velocity of the galaxy. Galaxies are involved in motions apart from the Hubble flow, such as falling onto a cluster under gravity, these result in a distortion of the perceived distance to a galaxy \citep{1987MNRAS.227....1K}. Because clusters grow under gravity, through studying their growth at different redshifts by isolating these redshift space distortions, we are able to learn how gravity behaves at different epochs. This may help us identify deviations from General Relativity \citep{2010PhRvD..81d3512S, 2009MNRAS.393..297P, 2005PhRvD..72d3529L}. It is well known that these two effects are degenerate in some regimes. Dynamic and geometric distortions can both have an enhancing effect on the galaxy power spectrum in the line of sight.

In order to identify and study these effects it is necessary to find features that are known to be the same physical size in the radial and transverse directions and the same size at all redshifts - a so called cosmic ruler. Unfortunately astrophysical objects are neither uniform nor large enough to perform such tests. The scale of the Baryon Acoustic Oscillations (BAO), i.e. the sound horizon at the epoch of last scattering, $r_s$, (seen on the CMB sky and first detected in the SDSS LRG galaxy distribution by \cite{2005ApJ...633..560E}) can be used as such a standard ruler to test our understanding of cosmology (for a recent review see \cite{2009arXiv0910.5224B}).  This characteristic clustering scale, which emerges from physical processes in the early universe, should be the same in both the radial and transverse directions.

In the transverse direction, measurement of the sound horizon gives a measurement of the angular diameter distance at which the sound horizon is observed, $d_A(z)/r_s$, where
\begin{equation}
d_A(z) = \frac{c}{H_0}\frac{r}{1+z},
\end{equation}
where
\begin{equation}
r(\chi) = \begin{cases}
\frac{\sin (\sqrt{-\Omega_k}\chi)}{\sqrt{|\Omega_k|}} \,\,\, \,\,& \Omega_k > 0\\
\chi \,\,\, \,\, & \Omega_k = 0\\
\frac{\sinh (\sqrt{\Omega_k}\chi)}{\sqrt{|\Omega_k|}} \,\,\, \,\,& \Omega_k < 0 \end{cases}
\end{equation}
$\chi(z)$ is the comoving distance
\begin{equation}
\chi(z) = H_0 \int_0^{z}\frac{dz'}{H(z')}.
\end{equation}
Thus the Hubble function and angular diameter distance are coupled in an FLRW universe. In the radial direction, the same BAO measurement gives $H(z)r_s$. Assuming that the radiation fraction is negligible the Hubble function is
\begin{equation}
H(z) = H_0\sqrt{\{\Omega_m(1+z)^3 + \Omega_k(1+z)^2 + \Omega_{DE} f(z)\}},
\end{equation}
where 
\begin{equation}
f(z) = (1+z)^{\{3(1+w_0+w_a))\exp(-3w_a z/(1+z)\}}.
\end{equation}
takes into account an evolving dark energy equation of state, $w$, which can be split into a constant, $w_0$, and time varying component, $w_a$,
 \begin{equation}
 w = w_0 + \frac{w_a}{1+z}.
 \end{equation}

There has been some debate regarding the detection of radial BAO. \citet{2009MNRAS.399.1663G} measured $H(z)$ using the BAO peak in the line of sight two-point correlation function. This was refuted by \citet{2010ApJ...710.1444K} who claimed it was consistent with background noise but used by others to get cosmological constraints \citep{2010arXiv1004.2599Z}. Criticisms of this detection were riposted by \cite{2011MNRAS.412L..98C}. Detections consistent with background noise can still be used to improve cosmological constraints if one assumes the $\Lambda$CDM cosmology. Future projects such as PAU\footnote{http://www.pausurvey.org}, DESpec\footnote{http://eag.fnal.gov/DESpec/Home.html} and BigBOSS\footnote{http://bigboss.lbl.gov} should provide stronger detections of the radial BAO.

Here we investigate how calibrating the BAO scale in the galaxy distribution to that measured in the CMB can remove the degeneracy between geometric and redshift space distortions improving constraints on cosmological parameters. Both of these probes, i.e. probes sensitive to the background expansion and the growth of perturbations, are important if one wants to distinguish the effect of dark energy from modified gravity.

Other similar work has been done on this by \cite{2010PhRvD..81d3512S}. They focused on looking at dark energy versus modified gravity and concluded that CMB calibration of the BAO scale was not a particularly powerful way of telling the difference between the two. Work by \cite{2011arXiv1102.1014S} is also similar, they focus on constraining curvature.

\cite{2010arXiv1004.4810S} make the interesting point that often when we consider constraints from combined probes we merely over plot the likelihood contours even though CMB information is routinely included in each probe. Meaning that it is included more than once in the final interpretation of the results. This highlights the need to understand CMB calibration.

We look at constraints from a  ``small" survey of volume $1(h^{-1}{\rm Gpc})^3$ and a ``large" survey of $20(h^{-1}{\rm Gpc})^3$. This large survey is exemplary of future spectroscopic galaxy redshift surveys like BigBOSS, a proposed all sky spectroscopic galaxy redshift survey and DESpec, a proposed spectroscopic follow up to the photometric Dark Energy Survey\footnote{http://www.darkenergysurvey.org/}. Surveys such as DESpec/BigBOSS will be large and deep enough to use redshift space distortions to measure the growth of structure in the same redshift bins used to measure the Hubble function from radial BAO measurements. We make the assumption that if redshift measurements are accurate enough to measure the radial BAO scale then the precision of these measurements will have little effect on our results (see \cite{2009ApJ...691..241B}). We do not, therefore, consider how uncertainties in redshift measurements will affect our results but instead leave this for further work. Nor do we consider non linear redshift space distortions.

In order to get a feel for the effect of CMB calibration and to avoid adding CMB information via a ``black box", we shall start our investigation with a toy model for the galaxy power spectrum (section \ref{toy model}).  This toy model includes the BAO and geometric and redshift distortions. Later we will investigate the problem with a numerically calculated power spectrum, produced by CAMB \citep{Lewis:1999bs} to allow for a time varying dark energy equation of state \citep{2008PhRvD..78h7303F} (section \ref{camb model}).

In section \ref{fisher section} we investigate the effect of CMB calibration on the parameters used to write our toy model using Fisher information. We show how a prior on just the sound horizon alone reduces the area of the Fisher matrix $1\sigma$ ellipse in the parameter space describing the geometric and redshift distortions.

In section \ref{Nested Sampling} we investigate how nested sampling explores the parameter space and show that derived constraints on parameters are not the same. Through the use of importance sampling (see section \ref{Nested Sampling} for an explanation) we produced combined constraints for our numerical galaxy power spectrum and CMB observations (using a publicly available WMAP chain \citep{Komatsu:2010fb}). We compare these constraints with those from the power spectrum and a prior on the sound horizon alone finding that they are not the same. We present our main results in section \ref{results} and discuss their implications in section \ref{Discussion}.

\section[A Toy Model for the Observed Galaxy Power Spectrum]{A Toy Model for the Observed Galaxy Power Spectrum}\label{toy model}

Here we construct a toy model for the observed galaxy power spectrum which includes BAO `wiggles' in addition to geometric and redshift space distortions. By design this toy model is not explicitly dependant upon cosmology acting as a fitting function for the observed anisotropic power spectrum. In reality, of course, the shape and distortions of the galaxy power spectrum are cosmologically dependant. The aim of this toy model is to focus clearly on quantities that are measured from the large scale structure (LSS) two point correlator. Our toy power spectrum is similar to that used by \cite{2010PhRvD..81d3512S} but simpler, excluding non-linear redshift space distortions.

The Bardeen, Bond, Kaiser, Szalay (BBKS) transfer function (equation \ref{BBKS}) gives an analytically simple power spectrum which is satisfactory for our purposes \citep{1986ApJ...304...15B}.
\begin{equation}\label{BBKS}
\begin{split}T(q) = \frac{\ln[1+2.34q]}{2.34q}
\times[1+3.89q + (16.2q)^2 \\+ (5.47q)^3 +(6.71q)^4]^{-0.25},\end{split}
\end{equation}
where $ q =\frac{k}{\Gamma}$, $\Gamma$, is the power spectrum shape parameter.
This gives us the linear, `no wiggles' part of our power spectrum.
\begin{equation}
P_{nb}(k) = A^*k^{n}T^2_{BBKS}(\frac{k}{\Gamma}).
\end{equation}
$A^*$ is the amplitude of the power spectrum and is independent of cosmology and $\Gamma$ is the shape parameter, which probes the matter density and baryon fraction thus \citep{1995ApJS..100..281S}:
\begin{equation}\label{Gamma}
\Gamma = \Omega_m h\exp\bigg(-\Omega_b - \sqrt{2h}\frac{\Omega_b}{\Omega_m}\bigg).
\end{equation}
$n$ is the spectral index, which for a Harrison-Zeldovich spectrum is equal to $1$ (inflation typically predicts that this number is less than $1$). The spectral index is different from the other parameters in this power spectrum since it can be considered a cosmological parameter in itself and not as such dependant on cosmology, we therefore keep it fixed at $n=1$.

An approximation to the baryon acoustic oscillations can be added to this by multiplying by a slowly varying sinusoidal function whose peaks are harmonics of the horizon scale $\sin(kr_s)$, with some damping given by some exponential part similar to the fitting function used by \cite{2005ApJ...631....1G}
\begin{equation}\label{pk}
P(k) = P_{nb}(k)\bigg(\frac{k}{\Gamma}\bigg)\bigg[1+A_{BAO}k\exp\bigg\{-\bigg(\frac{k}{k_s}\bigg)^{1.4}\bigg\}\sin(r_sk)\bigg].
\end{equation}

However, this is not how the power spectrum appears to the observer. We have to incorporate into our formalism descriptions of the linear dynamic and geometric distortions.  In redshift space, allowing for anisotropies, the power spectrum takes the form
\begin{equation}
P^z(k_\parallel,k_\bot) = \bigg[1+\beta(z)\frac{k_\parallel^2}{k_\parallel^2+k_\bot^2}\bigg]^2 P\bigg(\sqrt{k_\parallel^2 + k_\bot^2}\bigg),
\end{equation}
where
\begin{equation}\label{beta}
 \beta = \frac{\Omega_m^{\gamma(z)}}{b(z)},
\end{equation}
 $\Omega_m(z)^\gamma$ is the equivalent of the logarithmic rate of change of the linear growth factor with respect to redshift. $\gamma$ is dependant on the models of gravity and structure formation used and can be dependant on redshift. This simple parameterisation of the growth of structure has been shown to be effective in capturing the behaviour of a large class of models. Standard GR in a $\Lambda$CDM universe gives a constant $\gamma = 0.55$, whilst the DGP brainworld model is approximated by $\gamma = 0.68$ \citep{2007APh....28..481L}. Some theories of gravity require a $\gamma$ that slowly varies with redshift and in theories of gravity which allow for the presence of anisotropic stress $\gamma$ can be scale dependant. When designing experiments to measure $\gamma$ we will therefore have to consider measuring $\gamma$ in different redshift bins and at different scales in order to constrain deviations from GR. A simple extension to include a dependence on the dark energy equation of state is \citep{2005PhRvD..72d3529L} 
 \begin{equation}\label{little_gamma}
 \gamma(z) = 0.55 + 0.05(1+w(z)).
 \end{equation} 
 $b(z)$ is the bias, a measure of how well the galaxy distribution follows the underlying dark matter distribution. In reality this will be a function of redshift but for now we shall assume that it is a constant and is equal to unity. $k_\parallel$ and $k_\bot$ are respectively the parallel and perpendicular components of ${\rm k}$. 
 
 \cite{2011ApJ...727L...9J} show, using N body simulations, that strictly linear redshift distortions are very poor at measuring the growth rate in non $\Lambda$CDM cosmologies, even at large scales. However, we will follow the likes of \cite{2011arXiv1102.1014S} in assuming that linear redshift distortions are a good enough approximation at large scales.
 
Considering the Alcock-Paczynski effect we need to introduce parameters that describe the ratio between the actual and assumed cosmologies in the parallel and perpendicular directions,
\begin{equation}\label{c_par c_perp}
c_\parallel(z)=\frac{H^{fid}(z)}{H(z)}\,\,\,;\,\,\,c_\bot(z)=\frac{d_A(z)}{d_A^{fid}(z)}.
\end{equation}
These parameters relate to the Hubble parameter $H(z)$ and angular diameter distance $d_A(z)$. If our fiducial cosmology is correct (and therefore correctly describing our observed power spectrum) then $c_\parallel = c_\bot = 1$. Thus by measuring the sound horizon in both directions we can determine how well we know our cosmology. It should be noted that there is still a degeneracy between $H(z)$ and $d_A$ since both are dependant upon cosmology and are coupled. Meaning that a result of $c_\parallel \neq c_\bot \neq 1$ does not tell us \textit{how} our fiducial cosmology is wrong, merely that it is.

\begin{figure}
\centering
\psfrag{kpar}{$k_{\parallel}$}
\psfrag{kperp}{$k_{\bot}$}
\includegraphics[width = 8cm]{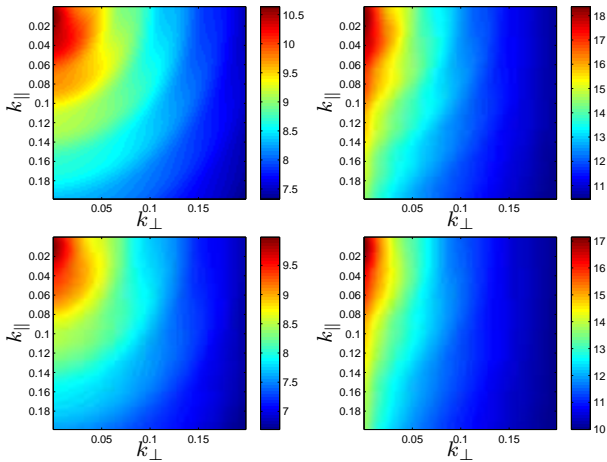}
\caption{The model power spectra used in our analysis, the colourbars have units of ${\rm ln}(h^-1 {\rm Mpc})$. The top left panel is the anisotropic toy model power spectrum (as outlined in section \ref{toy model}), the top right panel gives the error, $\sigma^2$, on this model, for the $20(h^{-1}{\rm Gpc})^3$ `DESpec/BigBOSS' survey with shot noise. The bottom two panels are the CAMB power spectrum (as outlined in section \ref{camb model}) and the error bars associated with it. The $k$bin size is $(1.9\times 10^{-3} h {\rm Mpc}^{-1})^3$. The colour scales used in these plots are logarithmic. }
\label{fig:pk_plots_print.eps}
\end{figure}

The observed power spectrum is then described by
\begin{equation}\label{obs_pk}
P^{obs}(k_\parallel, k_\bot;z) = \frac{1}{c_\parallel(z) c^2_\bot(z)}P^z(\tilde{k}_\parallel, \tilde{k}_\bot).
\end{equation}
Where $\tilde{k}_\parallel = k_\parallel / c_\parallel$ and $\tilde{k}_\bot = k_\bot / c_\bot$. The prefactor is needed because geometric distortions lead to a misestimation of the survey volume. The fiducial values for the parameters in this toy model are given in table \ref{base_parameters} and the power spectrum illustrated in figure \ref{fig:pk_plots_print.eps}.

\begin{center}
\begin{table}
\begin{tabular}{|c|c|c|c|c|c|c|}
\hline 
$A^*$ &$A_{BAO}$ &$\Gamma$ &$r_s$ &$\beta$ &$c_\parallel$ &$c_\bot$ \\
\hline
   3.80 &   2.00 &   0.15 &  149.86 &   0.52 &   1 &   1 \\
\hline
\end{tabular}
\caption{Fiducial values of the parameters of our toy power spectrum used to build the Fisher matrix (equations \ref{pk} - \ref{obs_pk}).}
\label{base_parameters}
\end{table}
\end{center}

\begin{center}
\begin{table}\label{camb_params}
\begin{tabular}{|c|c|c|c|c|c|c|c|}
\hline
$\Omega_{cdm}$ &$\Omega_b$ &$\Omega_{DE}$ &$w_0$ &$w_a$ &$H_0$ &$b$ &$A$ \\
\hline
   0.218 &   0.044 &   0.738 &  -0.827 &   -0.753 &  71.5 & 1.0 &   2.3 \\
\hline
\end{tabular}
\caption{Fiducial cosmology for our CAMB power spectrum, based on the best fit parameters for the WMAP with time varying dark energy equation of state.}
\label{fiducial_cosmology}
\end{table}
\end{center}

\section{A more realistic model for the galaxy power spectrum}\label{camb model}

The toy model, by design, is not explicitly dependant on cosmology. The distortion parameters can vary independently of the parameters that describe the shape of the power spectrum. In reality geometric and dynamic space distortions are dependant on the same cosmology that determines the shape of the power spectrum, so they cannot vary independently of the underlying power spectrum.

To build a more cosmologically based model we use a power spectrum generated by CAMB with geometric and redshift space distortions added in the same way as our analytic toy model. There are many parameters in CAMB that one could choose to vary. Here we have chosen parameters that can be related to those in our toy model. In particular, six of the parameters we have chosen determine the distortions on the power spectrum. The numerical model power spectrum is illustrated in figure \ref{fig:pk_plots_print.eps} and the fiducial values of its parameters given in table \ref{fiducial_cosmology}. 

In our toy model the amplitude, $A^*$, is a normalising factor, as is the amplitude parameter, $A$, that we vary in CAMB. The amplitude of the BAO, $A_{BAO}$, the power spectrum shape parameter, $\Gamma$ and sound horizon scale, $r_{\rm s}$, are dependant on the fraction of dark matter and of baryons and on $h$, so we vary $\Omega_{\rm cdm}$, $\Omega_{\rm b}$ and $H_0$. We assume a flat cosmology, so $\Omega_{\rm cdm} + \Omega_{\rm b} + \Omega_{\rm DE} = 1$ always. We also vary the linear bias parameter $b$ because the redshift space distortion parameter, $\beta$ is dependant on galaxy biasing. All other parameter values are kept fixed at the default CAMB values.

CMB data, such as that from WMAP or that which will come from Planck, contains more information than just the sound horizon scale and its error. It contains a plethora of cosmological information. Information contained in CMB observations on the geometry and the make up of the universe could also be used to improve parameter constraints. For example, the CMB gives us knowledge of $\Omega_{\rm m}$, this can be used as a prior on $\beta$. Indeed if we want to look for deviations from GR using measurements of $\beta$ it is first necessary to understand the matter fraction, $\Omega_{\rm m}$. In this paper we shall not investigate how explicit priors on parameters such as the matter fraction affect constraints, leaving this for further work. Instead we shall just investigate the prior on the sound horizon. 

\section{Fisher Analysis}\label{fisher section}

The Fisher Matrix formalism is a way of estimating the amount of information available in a given parameter space. It is a way of propagating errors in the case of multiple correlated measurements and many parameters. 

\cite{2009arXiv0906.0993B} or \cite{2009arXiv0901.0721A} provide a more detailed explanation of the statistics behind the Fisher matrix and also on how they are correctly manipulated. \cite{2007ApJ...665...14S} look at isolating the information cantained in the BAO peaks using Fisher matrices. We try to keep our use of Fisher matrices as simple as possible since later we will do a statistically more robust analysis.

The Fisher information is a measure of how much information an observable (in this case the power spectrum) contains about some unknown parameter $\theta$ (in this case the seven observable parameters in our toy model $A^*,A_{BAO},\Gamma,r_s,\beta,c_\bot,c_\parallel$). It is defined by the expectation value of the derivatives of the logarithm of the likelihood function, $\mathcal{L}$, with respect to parameters $\theta_i$.
\begin{equation}
F_{ij} = - \bigg\langle \frac{\partial^2 \ln \mathcal{L}}{\partial\theta_i \partial\theta_j} \bigg\rangle.
\end{equation}
Assuming that the noise on our data (the power spectrum) is Gaussian. The likelihood function for this observable can be expressed in terms of the theoretical value for that observable evaluated at some point, in our case at a particular $k$ value.

With some manipulation we can express the Fisher matrix as a sum over derivatives of the power spectrum with respect to parameters $\theta_i$ rather than as derivatives of the likelihood function.
\begin{equation}\label{fish_pk}
F_{ij}=\sum_n\frac{1}{\sigma_n^2}\frac{\partial P(k_n)}{\partial\theta_i}\frac{\partial P(k_n)}{\partial\theta_j}
\end{equation}
where $\sigma_n$ is the Gaussian error on $P(k_n)$. The error on the power spectrum is as follows.
\begin{equation}
\delta P = \sqrt{\frac{2}{4\pi k^2 \Delta k V}}\bigg(P + \frac{1}{\bar{n}}\bigg).
\end{equation}
Thus for the limit of large $\bar{n}$ i.e. ignoring shot noise only keeping the cosmic variance term
\begin{equation}
\delta P(k_n) = \sqrt{\frac{2}{4\pi k^2 \Delta k V}}P(k_n),
\end{equation}
where V is the volume of the survey. We will however always include shot noise in our calculations, assuming a constant number density of tracers $\bar{n} =  5\times10^{-4} (h^{-1} {\rm Mpc})^{-3}$. For a more detailed explanation see \cite{1998ApJ...499..555T}. (There is the implicit assumption that uncertainties in the measurement of the power spectrum are independent of cosmology. This assumption is incorrect since knowledge of the Hubble function and angular diameter distance are needed to convert redshifts to galaxy positions and derive a power spectrum from observational data.) We then put this into equation \ref{fish_pk}, $(\delta P)^2 = \sigma^2$, using $\partial P/ P = \partial \ln P$ and recognising that because the effects on the power spectrum that we are interested in are anisotropic we require the full two dimensional anisotropic Fisher matrix, we get:

\begin{equation}
F_{ij} = \sum_{k_\parallel,k_\bot}\frac{\partial \ln \tilde{P}(k_\parallel,k_\bot)}{\partial \theta_i}\frac{\partial\ln \tilde{P}(k_\parallel,k_\bot)}{\partial\theta_j} \frac{2\pi \Delta k^2_{\bot} \Delta k_\parallel}{(2\pi)^3}V.
\end{equation}
The logarithmic derivatives of the toy model power spectrum can be found in the appendix. We have assumed here that the different k-modes of the power spectrum are uncorrelated. In reality a given galaxy redshift survey will only cover part of the sky and may vary in depth. Thus different k-modes of the power spectrum are correlated by a window function and are not independant. The assumption of uncorrelated modes is valid if the $k$-bin $\Delta k >> 1/L$, where L is the smallest linear dimension of the survey, in our case this is $\sim 1 (h^{-1}Gpc)^3$. We used a $k$-bin size of $\Delta k = 1.9 \times 10^{-3} h {\rm Mpc}$ which satisfies this criterion.

In order to reduce the size of our parameter set we have two options. Either ``fix" the parameters we are uninterested in, i.e. assume that we know their values completely. If we think we know the value of a parameter completely then we can just remove that row and column from the Fisher matrix. This is equivalent to adding an infinite prior on that parameter. Or we can marginalise over that parameter. Treating it as a ``nuisance" parameter.

Constraints obtained from a marginalised Fisher matrix will always be weaker or equal to those from an unmarginalised matrix. Marginalisation will have no effect on the error on a parameter only if all other parameters are completely uncorrelated to it. This assumption breaks down because our observed parameters hide shared cosmological information and are therefore correlated. For example all our parameters except $A^*$ are dependant on $\Omega_{\rm m}$.

The procedure we follow to marginalise over unwanted parameters is to first invert the Fisher matrix to get the covariance matrix. Then we remove the row and column corresponding to the ``nuisance" parameter and reinvert to obtain the new reduced Fisher matrix. This is then repeated until we have marginalised over all our ``nuisance" parameters. 

In our toy model analysis we demonstrate calibration of LSS information to the CMB by the addition of a prior on the sound horizon scale $r_{\rm s}$. The sound horizon can be calculated using the description by \citet{1996ApJ...471..542H,1998ApJ...496..605E}. 

The prior matrix is simply a diagonal matrix made up of $Pr_{i=j} = 1/\sigma^2$, where $\sigma$ is the error on parameter $\theta_i$ and off diagonal components being zero. Adding this matrix to the original Fisher matrix then produces a matrix which includes the prior information. 

\begin{figure}
\psfrag{ampl}{$A$}
\psfrag{amplbao}{$A_{BAO}$}
\psfrag{cpar}{$c_{\parallel}$}
\psfrag{cpara}{$c_{\parallel}$}
\psfrag{cperp}{$c_{\bot}$}
\psfrag{beta}{$\beta$}
\psfrag{rs}{$r_s$}
\psfrag{shape}{$\Gamma$}
\centering
\includegraphics[width = 8cm]{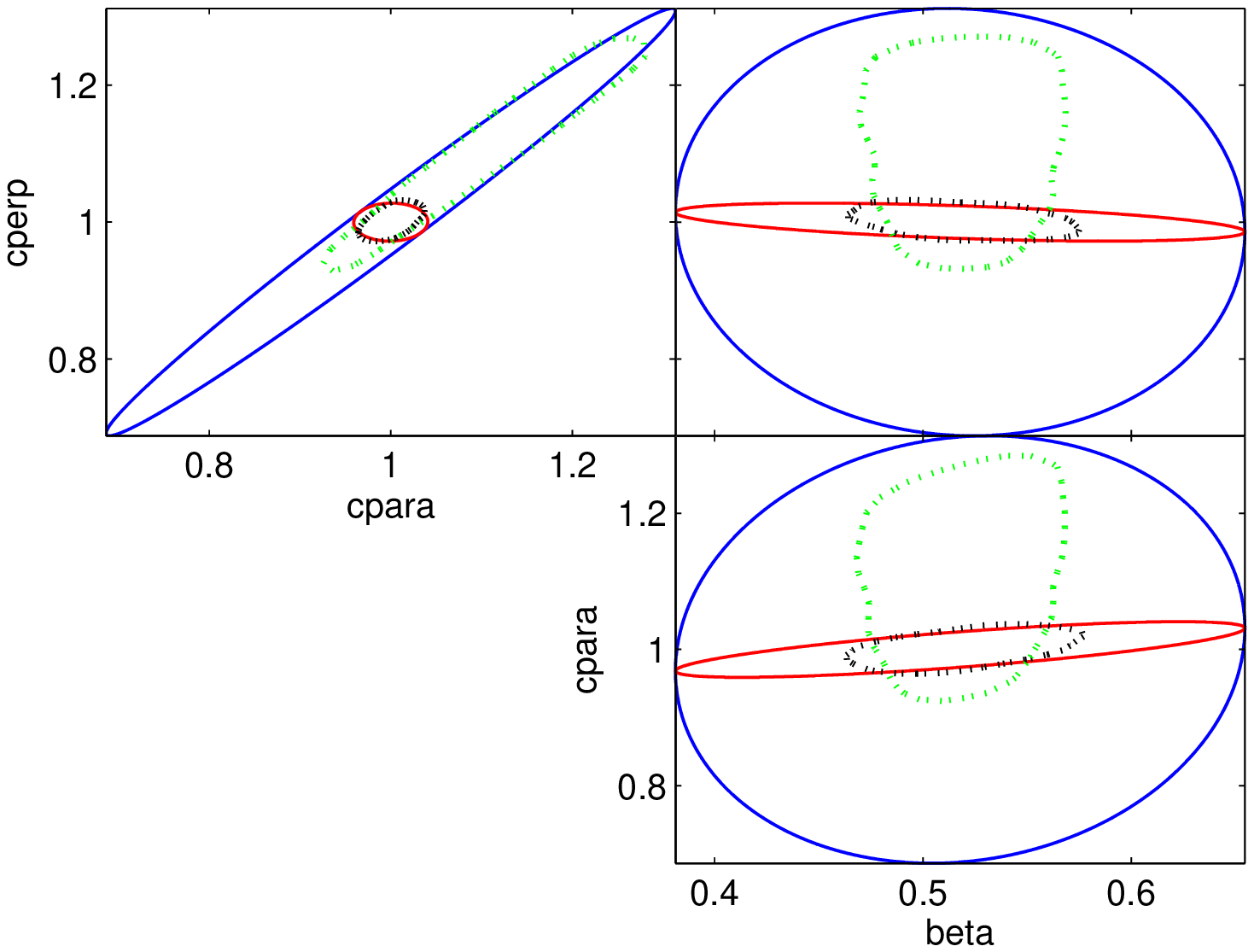}
\caption{An exploration of our toy model parameter space for a $1(h^{-1}{\rm Gpc})^3$ survey, focusing on the linear redshift distortion parameter $\beta$ and the line of sight and perpendicular geometric distortion parameters, $c_\parallel$ and $c_\bot$. Fisher matrix $1\sigma$ contours with (red) and without (blue) a prior on the sound horizon $r_{\rm s}$. Nested sampling $1\sigma$ contours with (black) and without (green) a prior on the sound horizon $r_{\rm s}$. Small galaxy redshift surveys are unable to estimate the values of cosmological parameters acurately alone. It is therefore difficult to approximate the errors on these paramters as being Gaussian.} 
\label{fig:toy_model_error_ellipses_LRG.eps}
\end{figure}

\begin{figure}
\psfrag{ampl}{$A$}
\psfrag{amplbao}{$A_{BAO}$}
\psfrag{cpar}{$c_{\parallel}$}
\psfrag{cpara}{$c_{\parallel}$}
\psfrag{cperp}{$c_{\bot}$}
\psfrag{beta}{$\beta$}
\psfrag{rs}{$r_{\rm s}$}
\psfrag{shape}{$\Gamma$}
\centering
\includegraphics[width = 8cm]{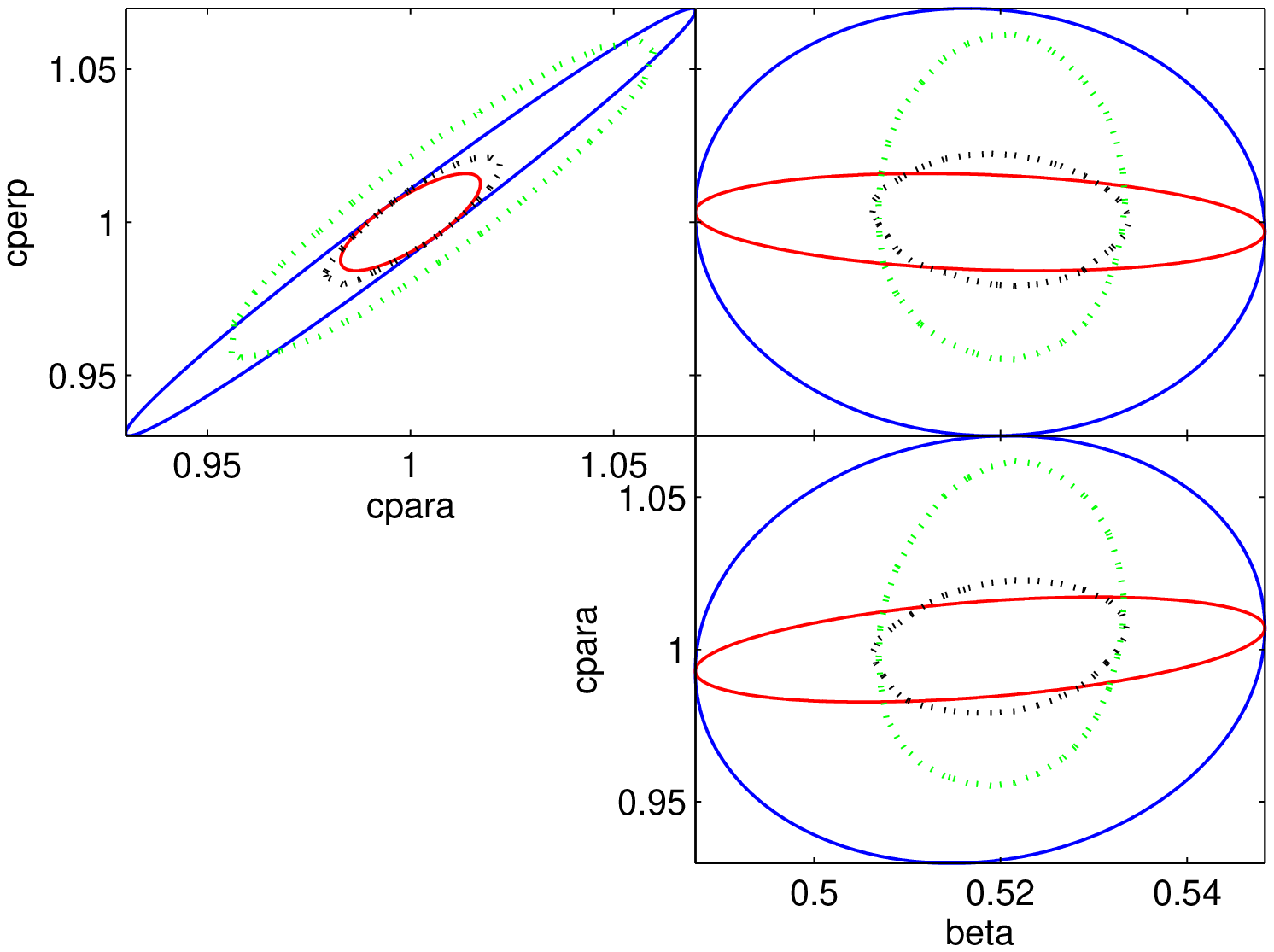}
\caption{An exploration of our toy model parameter space for a $20(h^{-1}{\rm Gpc})^3$ survey, focusing on the linear redshift distortion parameter $\beta$ and the line of sight and perpendicular geometric distortion parameters, $c_\parallel$ and $c_\bot$.  Fisher matrix $1\sigma$ contours with (red) and without (blue) a prior on the sound horizon $r_s$. Nested sampling $1\sigma$ contours with (black) and without (green) a prior on the sound horizon $r_{\rm s}$. }
\label{fig:toy_model_error_ellipses_DESpec.eps}
\end{figure}

\begin{figure}
\psfrag{cpar}{$c_{\parallel}$}
\psfrag{cperp}{$c_{\bot}$}
\psfrag{beta}{$\beta$}
\psfrag{rs}{$r_{\rm s}$}
\centering
\includegraphics[width = 8cm]{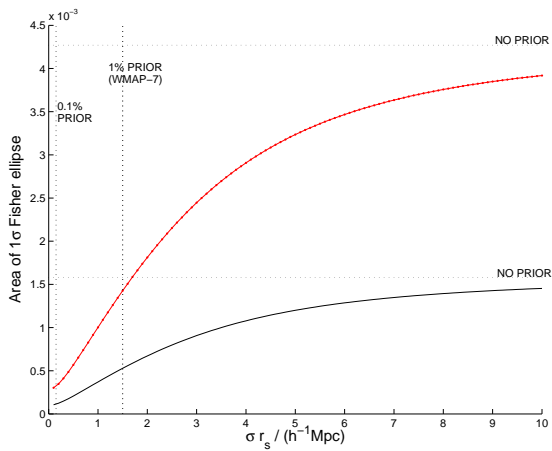}
\caption{Area of the 1$\sigma$ Fisher ellipse with respect to $\sigma_{r_s}$. Red line is $c_\parallel$-$\beta$, green dotted is $c_\bot$-$\beta$, blue dot dash is $c_\parallel$-$c_\bot$. The vertical black dotted lines show the value for $\sigma_{r_s}$ used as a prior in figures \ref{fig:toy_model_error_ellipses_LRG.eps} and \ref{fig:toy_model_error_ellipses_DESpec.eps} and a prior of $0.1 \%$. The blue dotted and green dotted lines show the area of $c_\bot-\beta$ and $c_\bot-c_\parallel$ ellipses respectively, with no prior on $r_{\rm s}$.}
\label{fig:ellipse_size_with_sigma_rs.eps}
\end{figure}

In Figures \ref{fig:toy_model_error_ellipses_LRG.eps} and \ref{fig:toy_model_error_ellipses_DESpec.eps} we have drawn error ellipses representing the constraints on pairs of parameters (marginalising over all other parameters) in our toy power spectrum from a $1(h^{-1} {\rm Gpc})^3$ volume limited survey (blue). The ellipses are then redrawn including only a prior on the sound horizon of $\sim 1 \%$, i.e. that $\sigma_{r_{\rm s}} = 1.4 \, {\rm Mpc} h^{-1}$ is a Gaussian error bar. This is about the size of the error bar on the the sound horizon scale from WMAP-7. The shrinking on the $c_\bot - c_\parallel$ ellipse implies that calibrating the sound horizon from the CMB to the sound horizon in the BAO features in the galaxy power spectrum can tell us whether we have the correct value for the angular diameter distance and Hubble function. The affect that decreasing the strength of the prior on the sound horizon has on constraints in the $c_\bot - c_\parallel$, $c_\bot - \beta$ and $c_\parallel - \beta$ planes is illustrated in figure \ref{fig:ellipse_size_with_sigma_rs.eps}. Note how calibrating the sound horizon has next to no effect of our constraints on $\beta$.

However, within $d_A(z)$ and $H(z)$ lie all the standard cosmological parameters (the dark energy equation of state appears in the Hubble function and in the angular diameter distance). In order to know how much information we have on those parameters from the original toy power spectrum we need to transform the original Fisher matrix. This is a straight forward coordinate transformation of a rank-2 tensor

\begin{equation}\label{transform}
\tilde{F}_{ij} = F_{rs}\frac{\partial\theta^r}{\partial\theta^i}\frac{\partial\theta^s}{\partial\theta^j}.
\end{equation}

When transforming to cosmological parameters including prior information from the CMB then becomes more complicated since many of the cosmological parameters we are interested in have priors from WMAP and other CMB experiments. They will need to be added to the cosmological parameters Fisher matrix before any further marginalisation if this information is to be included. We could do this by transforming an available Fisher matrix for a CMB probe into our base parameters and adding it to our Fisher matrix as in \citep{2010PhRvD..81d3512S}. However this does not help us \emph{understand} the effect of CMB calibration, it merely assumes it is a necessary step. Here we are trying to investigate the limit of information available before adding a prior from the CMB. To investigate cosmological parameter space we need a more physical toy model and more robust statistical techniques.

\section{Nested Sampling - a more robust exploration of parameter space}\label{Nested Sampling}

We are interested to see how our toy model with one Gaussian prior compares to a full likelihood analysis of a numerically produced power spectrum combined with real CMB information. One commonly used method in cosmology is   Markov-Chain Monte-Carlo (MCMC) by \cite{Lewis:2002ah}. Algorithms such as the Metropolis-Hastings algorithm attempt to optimally explore the parameter space without wasteful calculations at uninteresting points. 

Our chosen method of statistical parameter space exploration is `Nested Sampling' \citep{2004AIPC..735..395S,2006ApJ...638L..51M,2009MNRAS.398.1601F,2008MNRAS.384..449F}, a completely different algorithm to Metropolis-Hastings but with much the same aims.

It was first designed for Bayesian evidence evaluation, at which it is more efficient than MCMC since it transforms the multidimensional evidence integral into a one dimensional problem. This allows evidence to be evaluated from a single chain rather than multiple chains as is the case with traditional MCMC. Nested sampling can also be used for parameter estimation since it produces posterior inferences as a by product. Here we use the Multinest implementation of the nested sampling algorithm\footnote{http://ccpforge.cse.rl.ac.uk/gf/project/multinest/} \citep{2009MNRAS.398.1601F,2008MNRAS.384..449F}. There are other packages available, with different variations of the algorithm, such as Cosmonest \citep{2006ApJ...638L..51M}.

The likelihood for each link in the chain is calculated from the $\chi^2$ for that point and fed into the algorithm
\begin{eqnarray}
\chi^2 &=& \sum_{k_\bot, k_\parallel}\frac{(\tilde{P}^{fid}(k_\bot, k_\parallel) - \tilde{P}^{trial}(k_\bot, k_\parallel))^2}{\frac{2}{4\pi k_{\bot}^2\Delta k V} \bigg(\tilde{P}^{trial}(k_\bot, k_\parallel) + \frac{1}{\bar{n}}\bigg)^2},\\
\mathcal{L} &=& \exp{\frac{-\chi^2}{2}}.\label{pk_like}
\end{eqnarray}

We ran chains for  a $1 (h^{-1}{\rm Gpc})^3$ and a $20(h^{-1}{\rm Gpc})^3$ ``DESpec/BigBOSS" style survey, with shot noise $\bar{n} = 5 \times^{-4} {\rm Mpc}^{-1} h$ on our toy model power spectrum and on our numerically produced power spectrum.

CMB information was then included through importance sampling. Importance sampling enables us to combine information from different (independent) experiments (eg \citet{2010MNRAS.409.1100S}. Given a pre calculated Monte Carlo chain of parameter values $\theta$ which explores a likelihood distribution $L$, we can compute parameter constraints relative to a similar distribution $L'$ by reweighting the sample according to the likelihood ratio
\begin{equation}
w_{L'}^i = \frac{L'(\theta^i)}{L(\theta^i)}w_L^i
\end{equation}
For now we are interested in combining CMB and galaxy power spectrum information but we could combine any two independent experiments in this way. For two independent likelihood distributions,$L$ and $L'$, the combined likelihood distribution $\tilde{L}$ is just the product of the two
\begin{equation}
\tilde{L}(\theta)=L(\theta)L'(\theta).
\end{equation}
Which then allows us to simply reweight each point in the chain
\begin{equation}\label{reweight}
\tilde{w}^i = L'^i(\theta^i)w_L^i.
\end{equation}
Now likelihood contours drawn in the parameter space will represent the combined likelihood. It is worth noting that the `Metropolis Hastings' and `Nested sampling' algorithms use different weighting schemes, MCMC points have integer weights whilst the sum of the weights in a Multinest chain is one. For ease of comparison it is necessary to normalise the weights accordingly.

\begin{figure}
\psfrag{cpar}{$c_{\parallel}$}
\psfrag{cperp}{$c_{\bot}$}
\psfrag{beta}{$\beta$}
\centering
\includegraphics[width = 8cm]{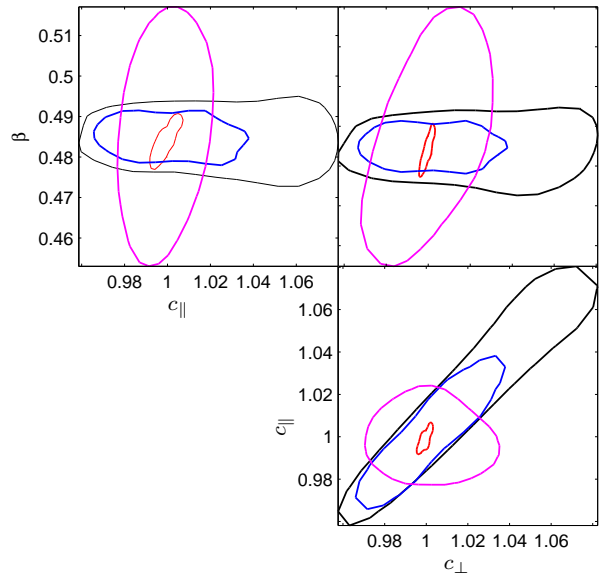}
\caption{Constraints on $\beta$ and $c_\bot$, $c_\parallel$ from nested sampling a CAMB power spectrum. The black ellipse represents the $1\sigma$ confidence limit based on a $20(h^{-1} {\rm Gpc})^3$ `DESpec/BigBOSS'-like survey alone. The purple contour is the $1\sigma$ confidence limit of the prior from WMAP projected onto this parameter space. The blue contours are the $1\sigma$ constraints from `DESpec/BigBOSS' importance sampled with the $1\%$ gaussian prior on $r_s$. The red contours are constraints from WMAP importance sampled with a `DESpec/BigBOSS' CAMB power spectrum projected onto this parameter space.}
\label{fig:all_impsamp.eps}
\end{figure}

We took a WMAP chain which allowed for a time varying dark energy equation of state and then calculated the likelihood that each point fitted the galaxy power spectrum using equation \ref{pk_like}. We then reweighted each point using equation \ref{reweight} and the WMAP weight. The Alcock-Pazcynski $c_\parallel, c_\bot$ and redshift space distortion $\beta$ parameters were then calculated for each point in the chain by comparing with the fiducial parameter values. This allows us to draw constraints in these planes (figure \ref{fig:all_impsamp.eps}).


To include a prior on the sound horizon in our toy model we assume the likelihood distribution around the true value of $r_s$ is a Gaussian of the same width as the prior used in the Fisher analysis. This is then importance sampled with the toy model, giving rise to the black contours in figures \ref{fig:toy_model_error_ellipses_LRG.eps} and \ref{fig:toy_model_error_ellipses_DESpec.eps}. To include the Gaussian prior in the CAMB model we calculated $r_s$ for each point in the chain using the prescription given in \citet{1996ApJ...471..542H,1998ApJ...496..605E}. We then drew a probability for this value from a Gaussian distribution peaked at the fiducial value for $r_s$ calculated using our fiducial parameter set in table 1, with the same standard deviation as before. Each point was then reweighted using equation \ref{reweight}, giving rise to the black contours in figure \ref{fig:all_impsamp.eps}. 

\section{Results}\label{results}

When comparing the $68 \%$ confidence limits in figure \ref{fig:toy_model_error_ellipses_LRG.eps} from the Fisher (blue) and nested sampling (green) of the toy model we can see that for smaller survey areas there is some tension between the two methods. The areas bounded by the contours are similar, with the exception of the constraints around $\beta$, but the two sets of contours do not cover exactly the same part of parameter space. The larger survey (figure \ref{fig:toy_model_error_ellipses_DESpec.eps}) is, however, in better agreement. Although in the $\beta-c_\parallel,c_\bot$ planes the areas covered by the contours still differ by about a factor of two.

\begin{figure}
\psfrag{ampl}{$A$}
\psfrag{amplbao}{$A_{BAO}$}
\psfrag{cpar}{$c_{\parallel}$}
\psfrag{cperp}{$c_{\bot}$}
\psfrag{beta}{$\beta$}
\psfrag{rs}{$r_s$}
\psfrag{shape}{$\Gamma$}
\centering
\includegraphics[width = 8cm]{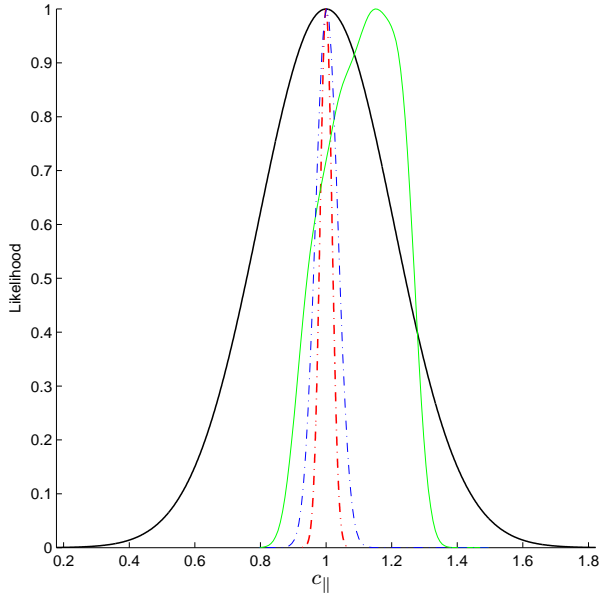}
\caption{Normalised probability distributions of $c_\parallel$ using our toy model for a $1(h^{-1} {\rm Gpc})^3$ survey with shot noise, $\bar{n} = 5\times10^{-4}(h^{-1} {\rm Mpc} )^{-3}$. The black solid line is the Gaussian probability distribution based on the Fisher matrix analysis. The solid green line is the Nested Sampling likelihood marginalised over all other parameters but with no prior on the sound horizon. The dot dashed blue line is the normalised 1d marginalised Gaussian probability distribution including a prior on the sound horizon. The dot dashed red line is the Nested Sampling likelihood marginalised over all other parameters and importance sampled with a Gaussian prior on the sound horizon.}
\label{fig:1dplot_LRG.eps}
\end{figure}

By design the maximum likelihood is the same in both cases, so the question arises, what is causing the disagreement? The answer can be found by examining the likelihood space in more detail. Let us look at the normalised 1d profile of the parallel Alcock-Paczynski parameter, $c_\parallel$. We can see that for a $1(h^{-1} {\rm Gpc})^3$ survey the likelihood is not Gaussian (figure \ref{fig:1dplot_LRG.eps}), being skewed. This puts asymmetric error bars on our parameters. When we look at the contours for a larger $20(h^{-1} {\rm Gpc})^3$ survey (the kind of volume expected for future spectroscopic surveys such as DESpec and BigBoss) we see that the spaces covered by the two sets of contours are more similar. If we look at the 1d probability distribution around $c_\parallel$ again we can see that it is more Gaussian than in the $1(h^{-1} {\rm Gpc})^3$ case since the marginalised probabilities and mean likelihoods are more similar. This is a pattern which is repeated for all the parameters in our toy model. The larger survey enables us to look closer into the peak of the distribution, which is more Gaussian. As anticipated, prior knowledge of the location and uncertainty in the sound horizon at last scattering improves our ability to constrain the shape of the power spectrum and geometric distortions.

\begin{figure}
\psfrag{ampl}{$A$}
\psfrag{amplbao}{$A_{BAO}$}
\psfrag{cpar}{$c_{\parallel}$}
\psfrag{cperp}{$c_{\bot}$}
\psfrag{beta}{$\beta$}
\psfrag{rs}{$r_s$}
\psfrag{shape}{$\Gamma$}
\centering
\includegraphics[width = 8cm]{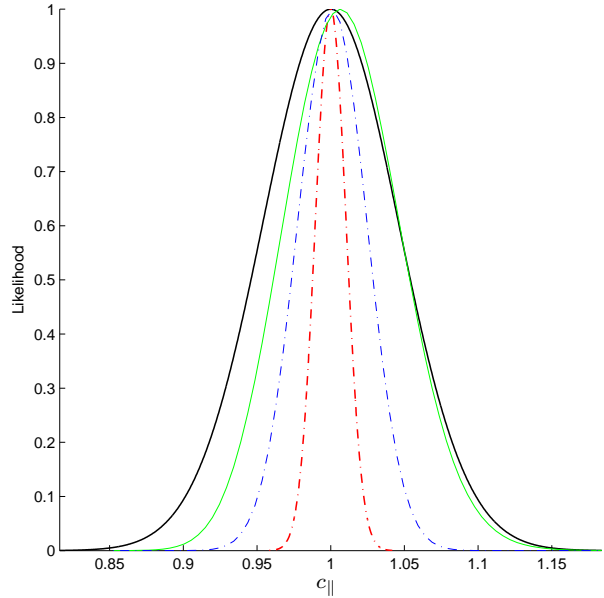}
\caption{Normalised 1d probability distributions of the line of sight geometric distortion parameter $c_\parallel$ using our toy model for a $20(h^{-1} {\rm Gpc})^3$ `DESpec/BigBOSS' survey with shot noise, $\bar{n} = 5\times10^{-4}(h^{-1} {\rm Mpc})^{-3}$. The black solid line is the Gaussian probability distribution based on the Fisher matrix analysis. The solid green line is the Nested Sampling likelihood marginalised over all other parameters but with no prior on the sound horizon. The dot dashed blue is the 1d marginalised Gaussian probability distribution including a prior on the sound horizon. The dot dashed red line is the Nested Sampling likelihood marginalised over all other parameters and importance sampled with a Gaussian prior on the sound horizon. The nested sampling probability distribution is more Gaussian than the $1(h^{-1} {\rm Gpc})^3$ case.}
\label{fig:1dplot_DESpec.eps}
\end{figure}

By adding a $~1 \%$ prior onto the sound horizon of $\sigma_{r_s} = 1.4 {\rm Mpc}\,h^{-1}$ in a Fisher matrix analysis we can see how calibrating with the CMB improves constraints in our $c_\bot - c_\parallel$ parameter space for a small $1(h^{-1}{\rm Gpc})^3$ survey by a factor of several and larger $20(h^{-1}{\rm Gpc})^3$ survey by more than a factor of two. 

We then conducted a likelihood analysis using nested sampling on our numerically produced power spectrum. A Gaussian prior on the sound horizon improves these constraints by a factor of two for the $20(h^{-1}{\rm Gpc})^3$, `DESpec/BigBOSS'-like survey. The combined constraints from the $20(h^{-1}{\rm Gpc})^3$ survey + WMAP are nearly a factor of 10 times better than the constraints from the survey alone. The WMAP prior contains considerably more information and is more orthogonal than the Gaussian prior on $r_s$.

\section{Discussion}\label{Discussion}

In this paper we have addressed two issues of great interest to current research in Cosmology, in particular in connection with the forthcoming generation of large galaxy redshift surveys. The first issue is physical - the callibration of the BAO using the sound horizon in the CMB. The second issue is statistical - a comparison of parameter fitting using Fisher matrix and Monte Carlo techniques.

We examined if the degeneracy between geometrical (Alcock-Paczynski) and dynamical (redshift space) distortions in the pattern of the galaxy distribution could be lifted. The key point here is to use the BAO in the pattern of the galaxy distribution as a feature of known physical size, the sound horizon $r_{\rm s} \approx 150 {\rm Mpc}$. We have shown, using a toy model, that this indeed can be callibrated using the equivalent feature in the CMB. By adding a prior onto the sound horizon of $\sim 1 \%$ we have shown, using a Fisher matrix analysis, error ellipses for line of sight and tangential distortion parameters shrink by \emph{by a factor of two} for a $20(h^{-1} {\rm Gpc})^3$ `DESpec/BigBOSS'-like survey with shot noise. This improvement is even more marked in smaller surveys, for $1(h^{-1} {\rm Gpc})^3$ the improvement is nearly a factor of 10. When our model for the power spectrum becomes more complicated (using CAMB) and we use the same Gaussian prior of $\sim 1 \%$ on the sound horizon the improvement of the ellipse is by a factor of $2.5$. When we start to include real data (from WMAP), which contains more information than just the sound horizon scale, then the improvement of the error ellipse is by a factor of $21$.

On the statistical side we have carried out a Monte Carlo Nested Sampling analysis on our parameter space. We find that Monte Carlo and Fisher methods can agree reasonably well for large volume ($\sim 20(h^{-1} {\rm Gpc})^3$) surveys with parameter constraints being nearly symetric and 68\% likelihood contours agreeing to within a factor of two. However, for smaller volume surveys the difference between constraints from the two methods can be greater with the areas of 68\% likelihood contours differing by more than a factor of four. The Nested Sampling analysis demonstrates that in general the likelihood distributions in the parameter space our toy model can be non-Gaussian. This undermines a central assumption in the Fisher matrix analysis, where errors are Gaussian by construction. All the likelihood distributions are strongly skewed, with the exception of that around $\beta$. This likelihood falls off more steeply than a Gaussian, having kurtosis. In this example better knowledge of $r_s$ does not help to improve constraints on $\beta$.

The Fisher constraints on the amplitude of the power spectrum for a $1 (h^{-1} {\rm Gpc})^3$ survey demonstrate the unphysicality of the Fisher formalism. Allowed within the $68 \%$ confidence limits are negative and zero values for $A^*$. Clearly in this case the Fisher information can only provide us with an upper limit. The Monte Carlo analysis \emph{does} provide us with an upper and a lower limit on the allowed values of $A^*$.

For most pairs of parameters the Monte Carlo and Fisher constraints are both markedly improved and more similar to each other with the addition of the prior on $r_s$. This is likely to be because the addition of a Gaussian prior makes the combined distribution more Gaussian, in the Monte Carlo case. There is no reason to assume, however, that CMB derived constraints on the sound horizon will be Gaussian.

The discrepancy between the Fisher likelihood space and perhaps more reliable Monte Carlo likelihood space is not a trivial point. 
When constraining the parameters of the toy model we allowed the parameters to vary independently but since the distortions are dependant on cosmology they cannot truly be varied independently of the shape of the underlying dark matter power spectrum, crucially in the context of this paper, this includes the shape and scale of the BAO.

There is a large difference between the improvement of constraints when a Gaussian prior on the sound horiazon is used and when we start to include real data (from WMAP). This could be because our parameterisation is different but is probably just because nature is more complex than a toy model. The CMB and also the real world galaxy power spectrum contain complexities and subtleties which are poorly understood.

Importance sampling relies on the two experiments having similar likelihoods, i.e. the chain used covers most of the likelihood volume in both experiments. If it is the case that the experiments provide highly orthogonal information then importance sampling is not appropriate because only the peak of the secondary distribution will be explored. This will give rise to erroneously tight constraints. This may be cause to doubt the constraints we have derived by importance sampling the WMAP chain. It may be the case that if the WMAP chain sampled a wider area then the combined constraints would be broader.

Even relatively fast Monte Carlo algorithms such as Nested Sampling are slower and more computationally demanding than Fisher matrix methods. Therefore the Fisher information remains a useful starting point when exploring the likelihoods of cosmological parameters. Further understanding of the affect of CMB callibration could be attained by expanding the cosmological parameter set used to describe geometrically and dynamically distorted large scale structure measurements.

While joint analysis of cosmological data sets can be carried out as a `black box' we have attempted here to gain insight and intuition into the interplay between large scale structure and CMB data, and to quantify the limitations of the Fisher matrix approach relative to Monte Carlo techniques. 
This approach can be extended further by more detailed modelling of e.g. non-linear redshift distortions and biasing, and on the statistical testing of different MCMC codes. Given the huge human and financial efforts invovled in the next generation of galaxy surveys it is crucial to \emph{understand} the physical effects and on the other hand to make sure that the statistics are robust. The method presented here can be directly applied to the next generation of spectroscopic surveys such as DESpec and BigBOSS combined with Planck.

\section*{Acknowledgements}
The code written to generate figures \ref{fig:toy_model_error_ellipses_LRG.eps} and \ref{fig:toy_model_error_ellipses_DESpec.eps} used some subroutines taken from Fisher4Cast \cite{2009arXiv0906.0993B}.  

We would like to thank the anonymous referee, Enrique Gaztanaga, Adi Nusser, Will Percival and Anais Rassat for helpful comments on the manuscript. O. Lahav and A. J. Hawken would also like to thank the Weizman Institute of Science, where part of this research was carried out. F.B. Abdalla recognises support from the Royal Society in the form of a URF and OL acknowledges a Royal Society Wolfson Research Merit Award.
\bibliographystyle{mn2e}
\bibliography{Calibrating_the_BAO_scale_using_the_CMB}
\bsp

\section*{Appendix}\label{Appendix}
Below are thge logarithmic derivatives used to build the Fisher matrix of the toy model of the observed power spectrum presented in section \ref{toy model}.

\begin{align}
\frac{\partial\ln\tilde{P}}{\partial A} &= \frac{1}{A}\\
\frac{\partial\ln\tilde{P}}{\partial A_{\rm BAO}} &= \frac{1}{A_{\rm BAO} + \frac{\exp((\frac{\tilde{k}}{k_s})^\gamma)}{k\sin(\tilde{k}r_s)}}\\
\frac{\partial\ln\tilde{P}}{\partial \Gamma} &= -2\frac{\tilde{k}}{\Gamma^2}\frac{T'}{T}\\
\frac{\partial\ln\tilde{P}}{\partial r_s}&= \frac{A_{\rm BAO}\tilde{k}^2\cos(\tilde{k}r_s)}{\exp((\frac{\tilde{k}}{k_s})^\gamma) + A_{\rm BAO}\tilde{k}\sin(\tilde{k}r_s)}\\
\frac{\partial\ln\tilde{P}}{\partial c_\bot} &= -\bigg[\frac{2}{1+(\frac{c_\parallel k_\bot}{c_\bot k_\parallel})^2} + \frac{2}{1+(\frac{c_\bot k_\parallel}{c_\parallel k_\bot})^2(1+\beta)} + \frac{k_\bot^2 P'(\tilde{k})}{c_\bot^2\tilde{k}P(\tilde{k})}\bigg]/c_\bot\\
\frac{\partial\ln\tilde{P}}{\partial c_\parallel} &= \bigg[-3 + \frac{2}{1+(\frac{c_\parallel k_\bot}{c_\bot k_\parallel})^2} + \frac{2}{1 + (\frac{c_\bot k_\parallel}{c_\parallel k_\bot})^2(1+\beta)} - \frac{k_\parallel^2 P'(\tilde{k})}{c_\parallel^2\tilde{k}P(\tilde{k})}\bigg]/c_\parallel\\
\frac{\partial\ln\tilde{P}}{\partial \beta} &= \frac{1}{1+\beta+(\frac{c_\parallel k_\bot}{c_\bot k_\parallel})^2}\\
\frac{\partial\ln\tilde{P}}{\partial n_s} &= \ln \tilde{k}
\end{align}

where $\tilde{k} = \sqrt{(k_\parallel/c_\parallel)^2 + (k_\bot/c_\bot)^2}$ and $P'$ is the first derivative of the power spectrum with respect to $k$.

\label{lastpage}
\end{document}